\documentstyle[useAMS,epsfig]{mn2e}
\def\etal{et al.}
\def\planck{{\sl Planck }}
\def\wmap{{\sl WMAP }}
\newcommand{\bi}[1]{\mbox{\boldmath $#1$}}

\def\alm{a_{\ell m}}
\def\Ylm{Y_{\ell m}}
\def\Cl{C_{\ell}}

\def\summ{\sum_{m=-\ell}^{\ell}}
\def\suml{\sum_{\ell=1}^{\infty}}

\title[Auto and cross correlation of phases of 1-year \textsl{WMAP}
  maps]{Auto and cross correlation of phases of the whole-sky CMB and
  foreground maps from the 1-year \textsl{WMAP} data}  

\author[Chiang \& Naselsky]{Lung-Yih Chiang$^{1}$\thanks{E-mail : {\tt
  chiang@nbi.dk}}, Pavel D. Naselsky$^{1,2}$ \\  
$^1$ Niels Bohr Institute, Blegdamsvej 17, DK-2100 Copenhagen,
Denmark \\ 
$^2$ Rostov State University, Zorge 5, 344090 Rostov-Don, Russia}

\date{Accepted 2005 ???? ???; Received 2005 ???? ???}

\begin{document}
\maketitle

\begin{abstract}
The issue of non-Gaussianity is not only related to distinguishing the
theories of the origin of primordial fluctuations, but also crucial
for the determination of cosmological parameters in the framework of
inflation paradigm. We present a method for testing
non-Gaussianity on the whole-sky CMB anisotropies. This method is based on
the Kuiper's statistic to probe the two-dimensional uniformity on a
periodic mapping square associating phases: return mapping of phases of the
derived CMB (similar to auto correlation) and cross
correlations between phases of the derived CMB and foregrounds. 
Since phases reflect morphology, detection of cross correlation of
phases signifies the contamination of foreground signals in the
derived CMB map. The advantage of this method is that one can cross
check the auto and cross correlation of phases of the derived maps and
foregrounds, and mark off those multipoles in which the
non-Gaussianity results from the foreground contaminations. We
apply this statistic on the derived signals from the 1-year WMAP
data. The auto-correlations of phases from the ILC map shows the
significance above 95\% CL against the random phase hypothesis on 17
spherical harmonic multipoles, among which some have pronounced cross
correlations with the foreground maps. We find that most of the
non-Gaussianity found in the derived maps are from foreground
contaminations. With this method we are better equipped to approach
the issue of non-Gaussianity of primordial origin for the upcoming
\planck mission.   
\end{abstract} 


\newcommand{\beq}{\begin{equation}}
\newcommand{\eeq}{\end{equation}}
\newcommand{\be}{\begin{eqnarray}}
\newcommand{\ee}{\end{eqnarray}}
\newcommand{\Odm}{\Omega_{\rm dm}}
\newcommand{\Ob}{\Omega_{\rm b}}
\newcommand{\Om}{\Omega_{\rm m}}
\newcommand{\nb}{n_{\rm b}}
\newcommand{\num}{\nu_\mu}
\newcommand{\nue}{\nu_e}
\newcommand{\nut}{\nu_\tau}
\newcommand{\nus}{\nu_s}
\newcommand{\mnus}{M_s}
\newcommand{\taus}{\tau_{\nu_s}}
\newcommand{\nnt}{n_{\nu_\tau}}
\newcommand{\rnt}{\rho_{\nu_\tau}}
\newcommand{\mnt}{m_{\nu_\tau}}
\newcommand{\tnt}{\tau_{\nu_\tau}}
\newcommand{\rar}{\rightarrow}
\newcommand{\lar}{\leftarrow}
\newcommand{\lrar}{\leftrightarrow}
\def\l{{\ell}}
\newcommand{\dm}{\Delta m}
\newcommand{\dl}{\Delta \l}
\newcommand{\lm}{{\l m}}
\def\kms{\ifmmode{{\rm km}\,{\rm s}^{-1}}\else{km\,s$^-1$}\fi}
\def\mpc{{\rm Mpc}}

\section{Introduction}
The issue of non-Gaussianity in the cosmic microwave background (CMB)
has touched the most fundamental base in cosmology. It was first brought to
attention by Ferreira, Magueijo and G\'{o}rski \shortcite{cobeng} that
non-Gaussian signal is present in the COBE data. Although it is almost
certain that the departure from Gaussianity is induced by systematic error
\cite{bandaycobeng,joaong}, the discussions and focus about
non-Gaussianity since then have been focusing primarily on primordial
origin. Mechanisms other than the simplest inflation model that have
been proposed for primordial density fluctuations produce non-Gaussian fields
(see Bartolo et al. 2004 and references therein). Due to this reason,
the issue of non-Gaussianity
seems to be discussed separately from that of the determination of
cosmological parameters. It is in the framework of inflation paradigm that the
cosmological parameters can only be determined correctly from
the angular power spectrum if the CMB temperature
anisotropies constitute a Gaussian random field (GRF). Therefore,
the issue of non-Gaussianity is not {\it beyond} the power spectrum,
but still {\it within} the power spectrum.       
   
The statistical characterization of temperature fluctuations of
CMB radiation on a sphere can be expressed as a sum over spherical
harmonics:
\begin{equation}
\Delta T(\theta,\varphi)=\suml \summ \alm \Ylm (\theta,\varphi),
\end{equation}
where $\alm=|\alm| \exp(i \phi_{\lm})$. The
strict definition of a homogeneous and isotropic GRF, as a result of
the inflation paradigm, requires that the moduli $ |\alm|$ are
Rayleigh distributed and the phases
$\phi_{\lm}$ are uniformly random on the interval $[0,2\pi]$. The
central limit theorem, however, guarantees that a superposition of a large
number of harmonic modes will be close to a Gaussian as long as the
phases are random. Hence the random-phase hypothesis on its own serves as a
definition of Gaussianity \cite{bbks,be}.

One of the most useful properties of GRF is that the second-order
statistics, the 2-point correlation function or the angular power
spectrum $\Cl$  
\begin{equation}
\langle  a^{}_{\ell^{ } m^{ }} a^{*}_{\ell^{'} m^{'}} \rangle = \Cl \;
\delta_{\ell^{ } \ell^{'}} \delta_{m^{} m^{'}} 
\end{equation}
furnish a complete description of the GRF. It is based on this
analytically-simple but important property that the cosmological
parameters can be correctly determined from $\Cl$. Accordingly, if
non-Gaussian signals are present, either with primordial origin, or
induced from data processing or systematic error, the
cosmological parameters derived from the $\Cl$ of such a
``contaminated'' field will have larger error bars. The issue of
non-Gaussianity in CMB is therefore not only
related to discriminating the theories of origin of primordial
fluctuations, it is also fundamental, in the framework of inflation
paradigm, for the determination of the cosmological parameters.

To test non-Gaussianity, the next order statistics: the 3-point correlation
function, or its Fourier transform, the bispectrum are often used. The
higher-order statistics, however, are only part of the whole
picture about non-Gaussianity. It takes a full hierarchy of $n$-point
correlation functions, or the polyspectra, to complete the statistical
characterisation of the CMB anisotropies.

Since the release of 1-year \wmap data, great efforts have been made for
the search and detection of non-Gaussianity via various approaches
and methods
\cite{wmaptacng,park,eriksenmf,santanderng,copi,eriksenasym,romanng,hansen,mukherjee,larson}.
These detection of non-Gaussianity shall augment the error bars on the estimation in cosmological parameters.

Practically speaking, characterizing non-Gaussianity through phases is
one of the most general approaches. Based on the random-phase hypothesis,
the key to developing statistical methods using phases is testing their
``randomness''. Due to the $2\pi$ wrapping, however, the phases tend to be
uniformly distributed at $[0, 2\pi]$. For example, for a point source
produces phases are distributed evenly and orderly, but not
randomly between 0 and $2\pi$. Probing non-Gaussianity via examining uniformity of phases themselves between 0 and
$2\pi$ is often as ineffective as via examining one-point Gaussian (temperature) probability
distribution $p(T)$ (e.g. it is possible $p(T)$ is still Gaussian for a non-Gaussian field). The $2\pi$ wrapping causing a uniform distribution of phases is similar to the central limit theorem in action producing one-point Gaussian probability distribution. We thus seek {\it associations} between phases as a more sensitive and effective statistical measure. 

The linear association such as auto correlation $\langle \phi
\phi^{'} \rangle$, however, does not give useful statistics due to 
the circular nature of phases. To counter this problem, return mapping of
phases is introduced to associate phase pairs systematically
\cite{phasemapping}. The main idea is mapping all phase pairs with the
same separation $(\dl,\dm)$ onto a square, which is conceptually
similar to the auto correlation 
$\xi(\dl, \dm)= \langle \phi_{\l,m} \,\,
\phi_{\l+\dl,m+\dm} \rangle$.

Another important feature of phases is that phases are closely related
to morphology. Through pixel-by-pixel cross correlation, maps with the
same phases display strong resemblance in
morphology, regardless of their power spectrum \cite{c3}. The level of
cross-correlation of phases taken from two images therefore renders 
significance of resemblance between them. Based on the
simple but prevailing assumption that the CMB
signals should not correlate with the foregrounds (i.e. the microwave
foregrounds should not have knowledge in what the CMB signals `look
like'), cross correlations of phases between the derived CMB and the
foregrounds shed light on the status of microwave foreground cleaning. Dineen and Coles
\shortcite{faraday} perform cross correlations in pixel domain between
the derived CMB and the foreground maps. Naselsky, Doroshkevich \&
Verkhodanov \shortcite{ndv03,ndv04}, Naselsky et
al.\shortcite{foreground} use cross correlation of phases to
illustrate the foreground contaminations in the derived CMB signals.

Return mapping of phases renders phase associations on a square with
each side ranged $[0, 2\pi]$ (with periodic boundaries). Cross correlation
of phases between derived CMB and foregrounds connect phases from
the same $\alm$ mode of the maps also produce mapping in a $[0, 2\pi]$
square. The null hypotheses for both cases: random phases for a
Gaussian CMB sky and no cross correlation between the CMB and the
foregrounds shall result in random points on $[0, 2\pi]$
squares. Statistics that are developed for a square taking into account
of the periodic boundaries can be implemented both on the return
mapping (auto correlation of
phases) of the derived CMB signal and on cross correlation of
phases between CMB and foregrounds. More importantly, through checking
both auto and cross correlations of phases from the derived maps, one
can gain insight into the issue of non-Gaussianity.

In this paper we test on each spherical harmonic multipole
$\l$ the auto and cross correlations, as well as the overall correlations. This paper is arranged as
follows. In Section 2 we recap the phase
mapping technique and introduce the Kuiper's statistics. We apply the
Kuiper's statistics in Section 3 both on the auto and cross correlation
of phases of the 1-year \wmap derived maps and connect the phase correlations with non-trivial whitened $\Delta T$ distribution in 1D Fourier composition. The conclusion and
discussions are in Section 4.

\section{The phase mapping technique and the Kuiper's statistics}

\subsection{The return mapping of phases and two-dimensional
  uniformity} \label{proj} 
Chiang, Coles \& Naselsky \shortcite{phasemapping} have introduced a technique
called return mapping of phases to render associations between phase pairs on a
square. The idea is borrowed from the return map in chaotic dynamics
\cite{returnmap}. The phase pairs are taken systematically and mapped
onto a $[0, 2\pi]$ square. In a return map of phase pairs with the
separation $(\dl, \dm)$, for example, the mapped points formed from
phase pairs have the
coordinates $(\phi_{\l,m} \,\, , \phi_{\l+\dl,m+\dm})$ for all
possible $\l$ and
$m$. Because of the circular nature of phases,
the $[0, 2\pi]$ return map is periodic at all 4 sides, which can be
viewed as a flat torus. 

\begin{figure}
\epsfig{file=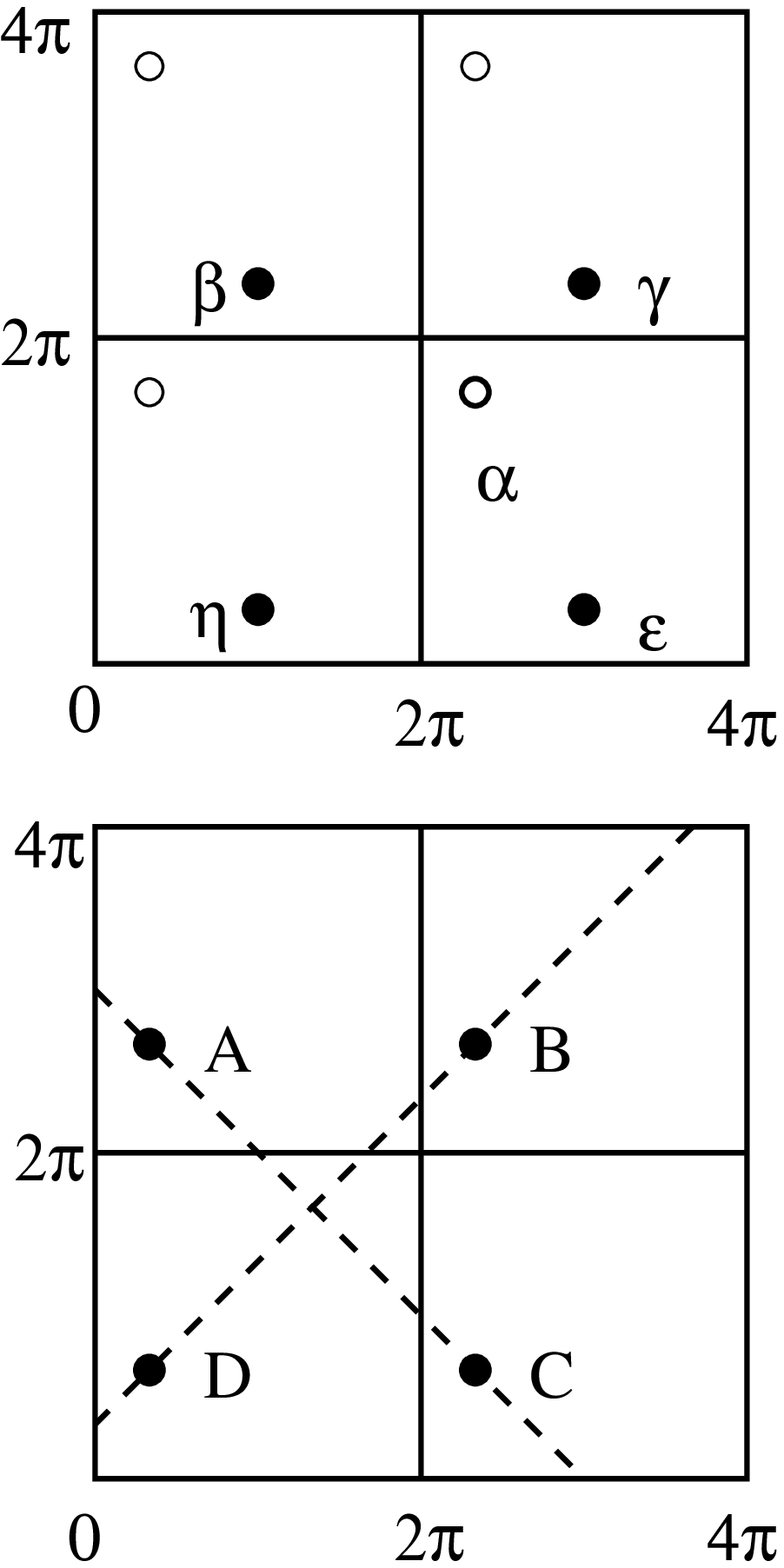,width=3.2cm}
\hspace{1cm}
\epsfig{file=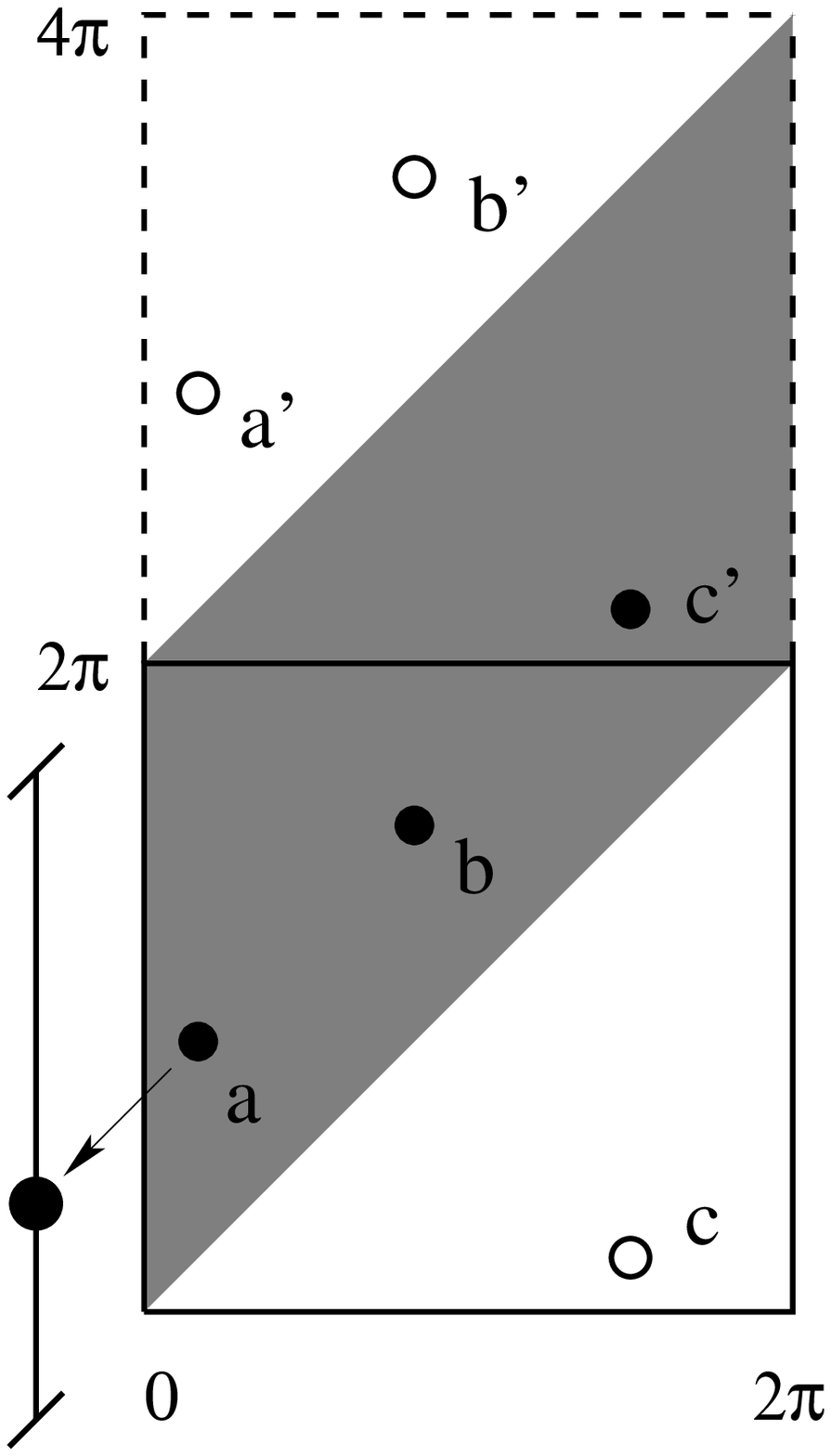,width=3.5cm}
\caption{We show two panels formed with a mosaic of 4 return maps of phases
  (left) and the projection of the mapping points to tackle the
  circular nature of phases. On the left top, the 4 points $\beta$,
  $\gamma$, $\varepsilon$ and $\eta$ on the return maps in fact
  represent the same point due to periodicity of the return maps. To
  probe the connectivity of points on the return
maps, therefore, we consider the simplest directions: anti-diagonal
  ($\vec{\rm BD}$), and diagonal ($\vec{\rm AC}$) projecting
  onto the $y$ axis. On the
  right panel, we put 3 points ${\bi a}$, ${\bi b}$ and ${\bi c}$, as
  an example on the return map and probe the connectivity in the
  anti-diagonal direction. We 
reproduce the same map on top of it: ${\bi a^{'}}$, ${\bi b^{'}}$ and
${\bi c^{'}}$. Due to the projection, we consider the same total area but
different configuration (the shaded area), which covers the 2 points
from the lower map (${\bi a}$ and ${\bi b}$), and 1 from the upper map
(${\bi c^{'}}$). The 3 seemingly
non-correlated points in a single return map now show the alignment
along the anti-diagonal direction.} \label{mosaicprojection}  
\end{figure}

After the mapping of phases, various statistics can be applied to
extract the statistical significance. Chiang, Naselsky \& Coles
\shortcite{meanchisquare} use a simple mean chi-square statistic to 
extract the information of the {\it one-dimensional} uniformity of the
distribution of the return maps. In order to
probe associations between points (uniformity in 2 dimensions),
care has to be taken on the periodicity of the return maps. In
Fig.\ref{mosaicprojection} on the left we show 
the mosaic of 4 return maps to indicate the periodicity. To probe the
connectivity of points on the return maps, we consider the
following 2 directions: anti-diagonal ($\vec{\rm BD}$), and
diagonal ($\vec{\rm AC}$) onto the $y$ axis. This projection is also
useful for cross correlation of phases. Complete cross correlation of
phases produces points exactly on the anti-diagonal line.

The projection of a point located at $(x_0, y_0)$ along the anti-diagonal
direction to $y$ axis as shown in Fig.\ref{mosaicprojection} is equivalent to
taking the difference of its $x$ and $y$ coordinates,
i.e. $y_0 - x_0$; in the diagonal direction $y_0 + x_0$. In a return map of
phases with separation $(\dl, \dm)$, points are formed with phase pairs 
$\phi_\lm$ and $\phi_{\l+ \dl, m+\dm}$, the projection then becomes
$\phi_{\l+ \dl, m+\dm} - \phi_\lm$: the phase difference. In Coles et
al. \shortcite{coleskuiper} they probe phase correlations by
investigating the uniformity of the neighbouring phase difference
$\phi_{\l,m+1} - \phi_\lm$. This is equivalent to testing the
uniformity of the projected points on a return map of the
separation $(\dl, \dm) = (0,1)$. Following this line of thought, the
analysis of the projection in either the vertical or the horizontal
directions is that of the randomness of $\phi_\lm$ themselves.

\begin{figure}
\epsfig{file=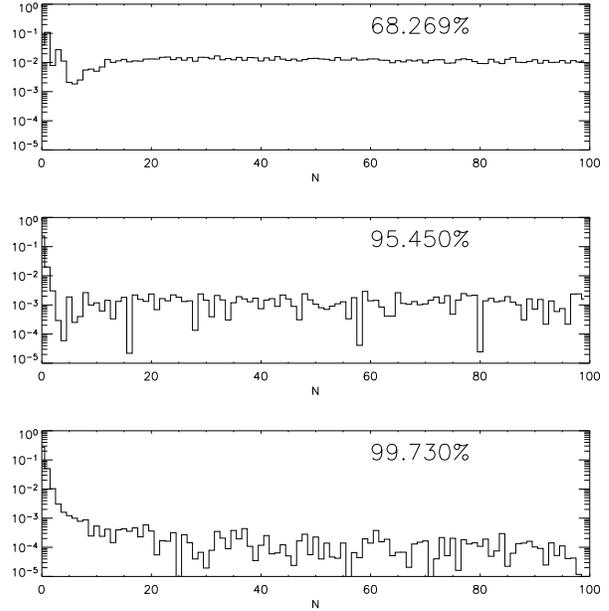,width=8.5cm}
\caption{The difference in the statistical significance of Kuiper statistics from its analytical derivation, Eq.(\ref{significance}) and (\ref{analytical}), and Monte Carlo simulation. We list the 3 frequently used significance value (from top to bottom): 66.269\%, 95.450\% and 99.730\%, respectively, which corresponds to Gaussian 1, 2, 3-$\sigma$ C.L.. The $y$-axis is the difference in absolute value, and the $x$-axis is the data point $N$.} 
\label{error}
\end{figure}

\subsection{The Kuiper's statistic}
The Kuiper's statistic can be viewed as a variant of
Kolmogorov-Smirnov (K-S) test \cite{kuiper,numerecipe,fisher}. As the
projection of points on the directions we mention in \S\,\ref{proj}
produces unbinned distribution that is a function of single
independent variable, it is very useful to apply the K-S test to probe
its uniformity. The K-S statistic is taken as the maximum distance of
the cumulative proability distribution against the theoretical one:  
\begin{equation}
D=\max_{-\infty < x < \infty} |S_N(x)-P(x)|
\end{equation}
For circular function, however, one needs to take into account of the maximum distance both above and below the $P(x)$     
\begin{eqnarray}
V=D_{+}+D_{-}& = &\max_{-\infty< x < \infty} [S_N(x)-P(x)] \nonumber \\
& + & \max_{-\infty<
  x < \infty} [P(x)-S_N(x)] \label{maxd}
\end{eqnarray}
and the C.L. against the null hypothesis (e.g. uniformity of phases in our case) can be calculated from 
\cite{numerecipe} 
\begin{equation}
{\rm C.L.}=1-Q_{\rm Kuiper}\left(V \left[ \sqrt{N}+0.155+\frac{0.24}{\sqrt{N}}
  \right] \right), \label{significance}
\end{equation}
where 
\begin{equation}
Q_{\rm Kuiper}(\lambda)=2\sum^{\infty}_{j=1}(4 j^2 \lambda^2-1)
e^{-2j^2\lambda^2},  \label{analytical}
\end{equation}
and $N$ is the data points. 
\begin{figure}
\epsfig{file=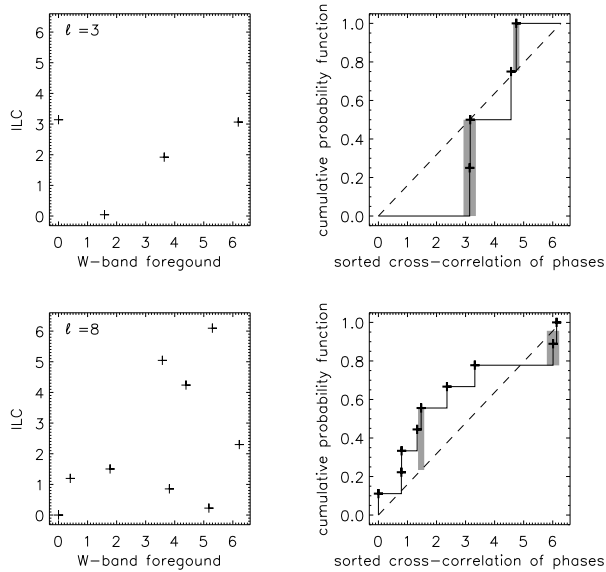,width=8.5cm}
\caption{The cross correlation of phases at $\l=3$ (top-left: $m=0-3$) and 8
(bottom-left: $m=0-8$) between the ILC map and the W-band foreground
map. One can imagine that, for example, for $\l=3$ the 4 points are
highly correlated along the anti-diagonal direction if the left-most
point is repositioned at the far-right. The same for $\l=8$ if we
reposition the 3 points at the lower-right region to the above top axis
line (cf Fig.\ref{mosaicprojection}). The panels on the right are their
corresponding cumulative probability functions. The dashed line is
the $P(x)$, the theoretical cumulative probability distribution for
2D uniformity. The thin and thick shaded lines are the $D_{+}$ and
$D_{-}$, respectively, their maximum distances of the cumulative
probability distribution above and below the theoretical one.}  \label{xcorr} 
\end{figure}

We run Monte Carlo (MC) simulation in order to test Eq.(\ref{significance}) and (\ref{analytical}). One hundred thousand realizations of the difference of two random phase series (the phases are defined $[0, 2\pi)$) are simulated, which is to mimic the projection of return map shown in Fig.\ref{mosaicprojection}. 
For each $N$, we have an ensemble of 100 000 random realization each producing maximum distance $V$ as in Eq.(\ref{maxd}). We then sort these 100 000 $V$ to yield the statistical significance from MC simulation. Simultaneously, the statistical significance can be calculated directly from Eq.(\ref{significance}) and (\ref{analytical}). 
In Fig.\ref{error} we show the difference in the statistical significance of the Kuiper statistics from its analytical derivation, Eq.(\ref{significance}) and (\ref{analytical}), and MC simulation. We list the 3 frequently quoted significance value (from top to bottom): 66.269\%, 95.450\% and 99.730\%, which corresponds to Gaussian 1, 2, 3-$\sigma$ C.L., respectively. One can see that for 95.450\% and 99.730\% the difference between analytical calculation and MC simulation is down to $10^{-3}$ level even for as few data points as $N=10$. 
 
We use the maps available on the WWW for testing this method: the derived foreground
maps at 5 frequency bands from the \wmap website \footnote{{\tt
http://lambda.gsfc.nasa.gov}} and the 4 derived CMB maps
: the internal linear combination (ILC) map by the \wmap science team
\cite{wmap,wmapresults,wmapfg,wmapsys,wmapcl,wmapng}, the ILC map by Eriksen \etal (hereafter EILC) \shortcite{eilc} \footnote{{\tt http://www.astro.uio.no/$\sim$hke/cmbdata/}}, the Wiener-filtered map (WFM) \footnote{{\tt
    http://www.hep.upenn.edu/$\sim$max/wmap.html}} by Tegmark, de
Oliveira-Costa \& Hamilton \shortcite{toh}, and the Phase-Cleaned Map (PCM)
by Naselsky \etal \footnote{{\tt http://www.nbi.dk/$\sim$chiang/wmap.html}} \shortcite{pcm}. Note that these maps are used only for testing the effectiveness of the Kuiper's statistics because of their different approaches on foreground cleaning and their different morphology.  Our claim about non-Gaussianity in this paper shall not contradict with others in the literature as most of these maps, as advised by the authors, are not for scientific purposes.

\begin{figure}
\epsfig{file=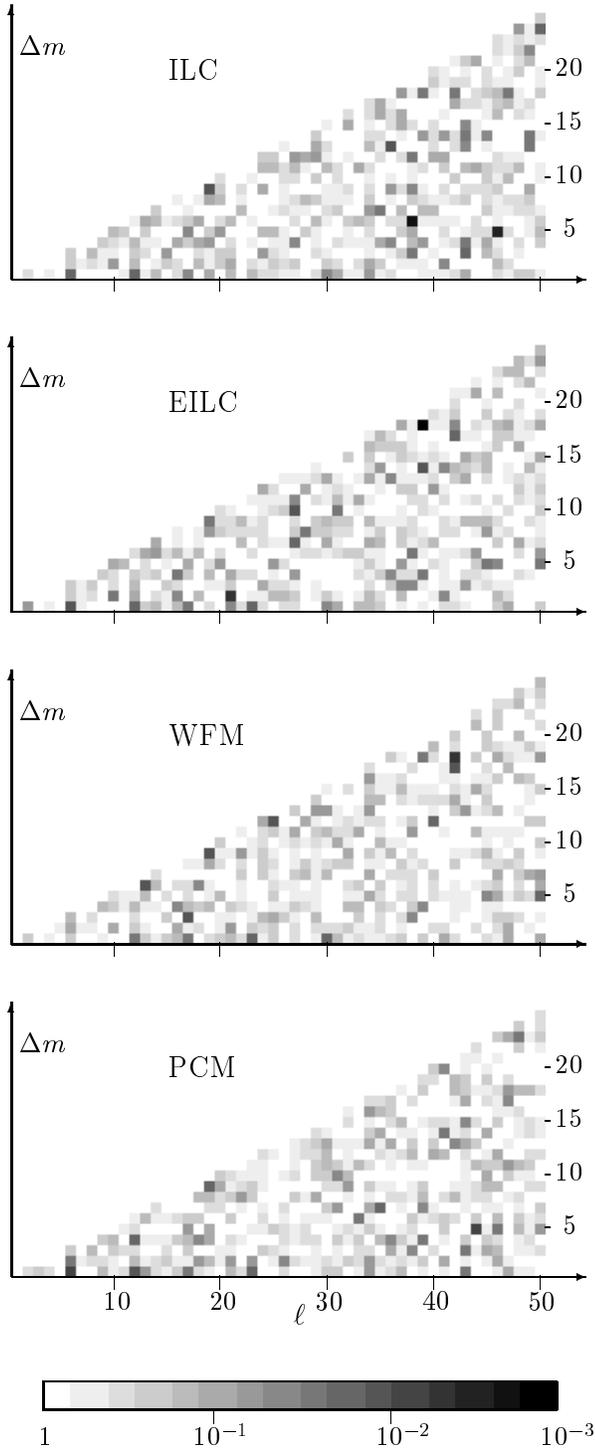,width=9cm}
\caption{The Kuiper's statistics on return maps for the \wmap ILC map, the EILC map, TOH's
Wiener-filter map and the PCM map. In order to display the C.L. more clearly against random phase hypothesis, we show the $Q_{\rm Kuiper}$ instead. So $10^{-3}$ corresponds to C.L. 99.9\% against uniformity of phases, $10^{-2}$ corresponds to C.L. 99\% \ldots etc.. Within each multipole $\l$ we take the phases of fixed separation $\Delta m$ and map them in a return map. We take $\Delta m=1$ up to $\Delta m=\l_0/2$ for multipole number $\l_0$. } \label{ksmapping}
\end{figure}
\begin{figure}
\epsfig{file=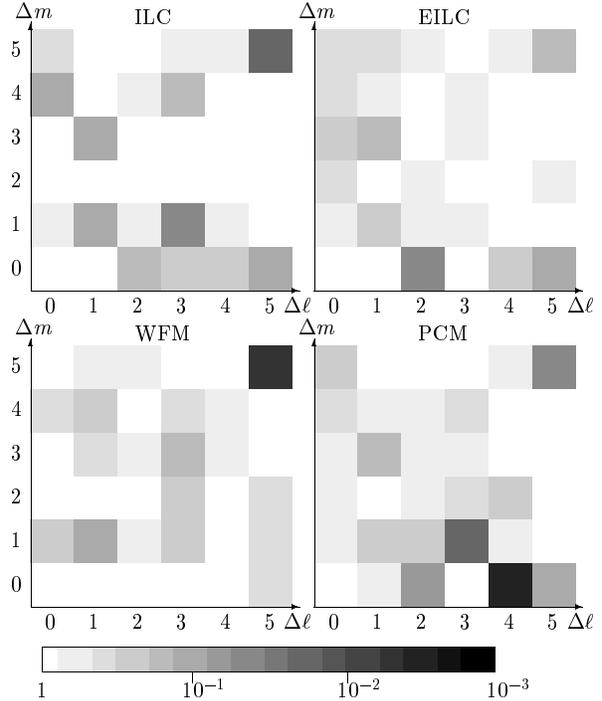,width=8.5cm}
\caption{The Kuiper's statistics on return map for separation $(\dl, \dm)$. The mapping takes all the phases from $\l=2$ to 50. The separation of mapping is ranged 0 to 5 for both $\delta \l$ and $\delta m$. In order to display more clearly the C.L. against random phase hypothesis, we show the $Q_{\rm Kuiper}$ instead. So $10^{-3}$ corresponds to 99.9\% C.L. against uniformity of phases, $10^{-2}$ to 99\% C.L. \ldots etc..} \label{mappingall}
\end{figure}
In Fig.\ref{xcorr} as an example of how the Kuiper's statistics works, we show the
cross correlation of phases between the ILC and the W-band foreground
map at $\l=3$ (top) and $\l=8$ (bottom) and
their corresponding cumulative probability functions from the
projection in the anti-diagonal direction onto the $y$ axis. We show with
thin ($D_{+}$) and thick ($D_{-}$) shaded lines the maximum distances
of the cumulative probability distribution above and below the
theoretical one, respectively.

\section{Auto and cross correlation of phases}

\subsection{Auto correlation}
We consider first within each multipole number $\l$ the mapping of phases of fixed $\dm$. In Fig.\ref{ksmapping} we show the Kuiper's statistics on return maps for the \wmap ILC map, the EILC map, the WFM and the PCM. Within each multipole $\l$ we take the phases of fixed separation $\dm$ and map them into a return map. For the multipole number $\l=\l_0$ we 
take the separation from $\dm=1$ to $\dm= \l_0/2$, although in general we can take any $\dm$ values to test the randomness of phases. In order to display the C.L. more clearly against random phase hypothesis, we show the $Q_{\rm Kuiper}$ instead. So $10^{-3}$ corresponds to C.L. 99.9\% against uniformity of phases, $10^{-2}$ corresponds to C.L. 99\% \ldots etc..

One interesting result is that for $\l=6$ the mapping of $\dm=1-3$ for all 4 maps are all above 75\% C.L.. and the
multipole numbers which has $\dm$ mappings that are above 95.45\%
C.L. against random phase hypothesis are $\l=6$, 12, 15,
17, 19, 30, $35-36$, $38-39$, 41, 43, $45-47$, $49-50$. For
$\dm=6$ at $\l=38$ and  $\dm=5$ at $\l=46$, both mappings reach above 99.73\%. For other 3 maps, they all have modes that reach 95.45\%. We do not, however, claim the overall non-Gaussianity of these maps.

We also consider the mapping for separation $(\dl, \dm)$ from all the available phases from $\l=2$ to 50 (except those of $m=0$ modes). In Fig.\ref{mappingall} we show the Kuiper's statistics for the 4 maps: the ILC, EILC, WFM and PCM. The separation of return mapping ranges from 0 to 5 for both $\dl$ ($x$ axis), $\dm$ $y$ axis), although it can be extended to higher number. The ILC, EILC and PCM have $(\dl, \dm)=(2,0)$ correlation above 68.27\% C.L. and there are considerable degrees of phase correlation at separation $(\dl, 0)$ for ILC, EILC and PCM, but except for WFM.  

Note also the strong correlation of $(\dl, \dm)=(4,0)$ for the PCM. This specific correlation is recently mentioned in Naselsky \& Novikov \shortcite{4nperiodicity}, which is caused by the symmetric (w.r.t. Galactic centre) point-like peaks lying along the Galactic plane. The PCM is produced without further cleaning performed for ILC, EILC and WFM, hence the Galactic contamination is still present.

\begin{figure*}
\epsfig{file=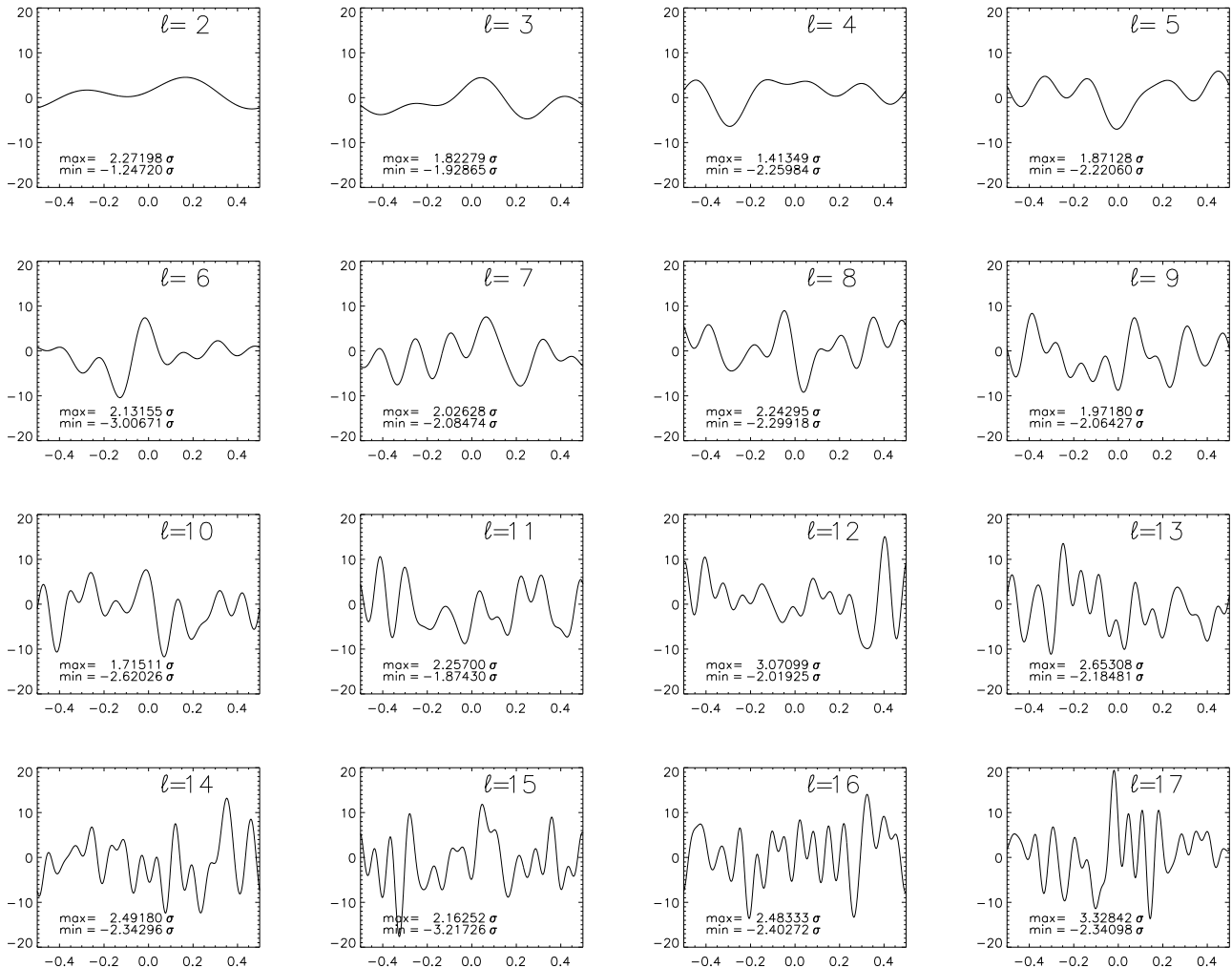,width=18cm}
\caption{The assembled ``whitened'' distribution $\l$ by $\l$ from the \wmap ILC map. We assemble only the phase part $\exp(i \phi_{\lm})$ of each harmonic number $\l$ in Fourier composition: $\Delta T_{\l,{\rm whitened}} (x)=\sum_m \exp(i \phi_{\lm}) \exp(i m x)$. The extrema that reach $3\sigma$ are $\l=6$, 12, 15 and 17: $-3.0\sigma$, $3.1\sigma$, $-3.2\sigma$ and $3.3\sigma$, respectively. These multiple numbers are picked up by phase correlations with C.L. 95.45\% from the Kuiper's statistics.}  \label{alm2k}
\end{figure*}
 
In order to see how non-random phases affect the distribution, we assemble only the phase part  in each harmonic number $\l$ in Fourier composition (the so-called ``whitened'' image): 
\begin{eqnarray}
\Delta T_{\l,{\rm whitened}} (x)& \equiv &\sum_m \frac{\alm}{|\alm|} \exp(i m x) \nonumber \\
 &=&\sum_m \exp(i \phi_{\lm}) \exp(i m x). \label{eq:assemble}
\end{eqnarray}
 
In Fig.\ref{alm2k} we display for $\l=2$ to 17 the ``whitened'' distributions which have no influence from the amplitudes $|\alm|$. This representation in 1D Fourier composition for each $\l$, rather than in the standard whole-sky spherical harmonic composition, can provide some advantages for further analysis. The morphology can be seen more clearly in 1D Fourier composition than in spherical harmonic composition. Here we can relate phase correlation (in Fourier space) to the simple GRF peak statistics (in real space). The extrema (maximum or minimum) that have reached $3\sigma$ appear in the distribution of $\l=6$, 12, 15 and 17: $-3.0\sigma$, $3.1\sigma$, $-3.2\sigma$ and $3.3\sigma$, respectively. These multipole numbers correspond to those picked up by the Kuiper's statistics with non-random phases at 95.45\% C.L.. Furthermore, phase correlation does not only manifest itself in the extrema. One can see, for example, for $\l=10$ the assembled whitened $\Delta T$ has repeated peaks. The C.L. against random phase hypothesis at $\dm=2$ of $\l=10$ reaches 94.13\%.

\subsection{Cross correlation}
The Kuiper's statistics can also be applied to cross correlation.  

In Fig.\ref{cl} we show the C.L. against the null hypothesis in
cross-correlation of phases for $\l=2$ to 50. The thick lines are the \wmap ILC
cross the \wmap foreground maps at (from top to bottom) K, Ka, Q, V and W-band, the cross sign ($\times$), the diamond sign ($\diamond$), the plus sign ($+$) are the EILC by Eriksen \etal \shortcite{eilc}, the PCM by Naselsky \etal \shortcite{pcm} and the WFM by Tegmark \etal \shortcite{toh}, respectively, cross the \wmap foreground maps. We would like to point out that the ILC are produced by less weights from the W-band map than V and Q.   
One can see that the ILC map has substantial cross correlation of phases for $\l=3$ and 8 with \wmap all 5 channel foreground maps. High cross correlation with the foregrounds, however, does not necessarily imply the non-Gaussianity for that multipole number.  

\begin{figure}
\epsfig{file=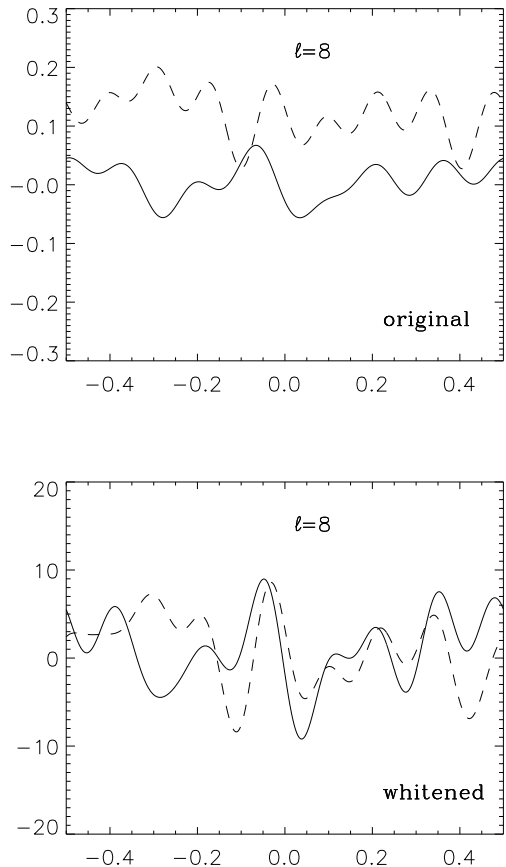,width=8cm}
\caption{The assembled $\Delta T$ distribution for $\l=8$ from \wmap ILC map (solid curve) and \wmap W-channel foreground map (dashed curve). The top panel is assembled from $\sum_m a_{8m} \exp(i m x)$ and the bottom (whitened) from $\sum_m \exp(i \phi_{8m}) \exp(i m x)$. Notice the resemblance in morphology between the derived CMB and foreground signals.} \label{assemble8}
\end{figure}

In Fig.\ref{assemble8} we show the assembled $\Delta T$ distribution for $\l=8$ from \wmap ILC map and \wmap W-channel foreground map. From the Kuiper's statistics for cross correlation of phases, for $\l=8$ the correlation reaches C.L. 89.98\%, which reflects on their resemblance in morphology. 
 
\section{Conclusion and Discussions} 
In this paper we introduce the Kuiper's statistic to probe the 2D uniformity for the phase mapping technique. The Kuiper's statistics are useful to explore both auto correlation and cross correlation of phases. We use the 4 maps to test the effectiveness of this method: \wmap ILC, the ILC by Eriksen \etal, the WFM by Tegmark \etal and the PCM by Naselsky \etal and found several multipole numbers with non-randomness of phases over 95.45\%. The Kuiper's statistics can also used to test on all the available phases.

Contrary to the $\Delta T$ representation for each $\l$ from standard spherical harmonic composition, we use 1D Fourier composition to display the $\Delta T$. We use the ``whitened'' Fourier composition to display the connection between non-random phases and non-trivial morphology, and between cross-correlated phases of 2 maps and resemblance in their morphology.   

\begin{figure}
\epsfig{file=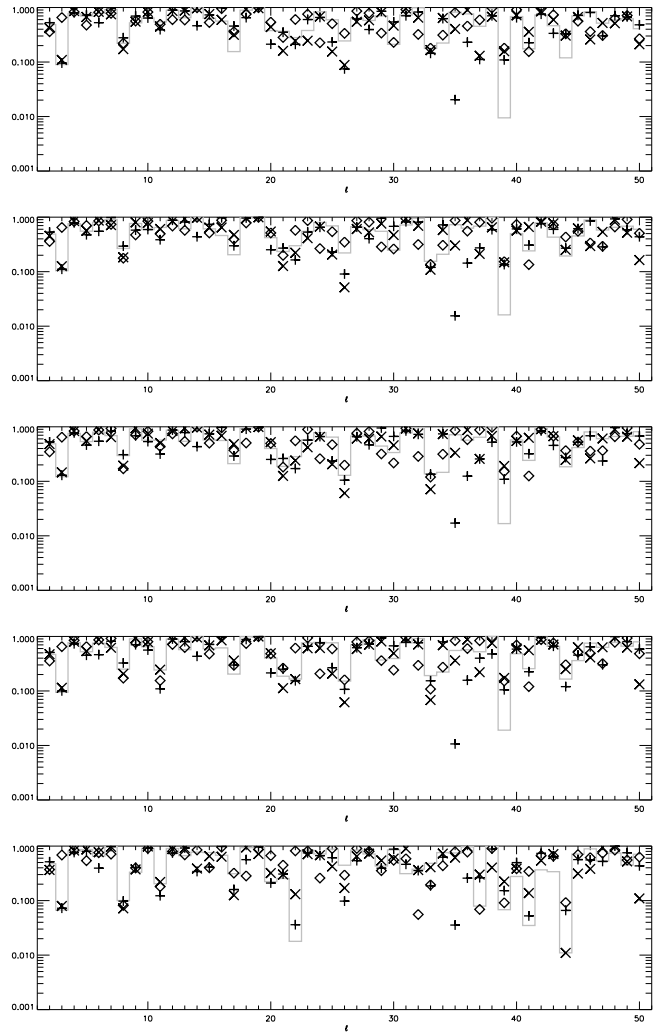,width=9cm}
\caption{The confidence level against the null hypothesis in
cross-correlation of phases for $\l=2$ to 50. The thick lines are the \wmap ILC
cross the \wmap foreground maps at (from top to bottom) K, Ka, Q, V and W-band, the cross sign ($\times$), the diamond sign ($\diamond$), the plus sign ($+$) are the EILC by Eriksen \etal \shortcite{eilc}, the PCM by Naselsky \etal \shortcite{pcm} and the WFM by Tegmark \etal \shortcite{toh}, respectively, cross the \wmap foreground maps. In order to display the C.L. against random phase hypothesis more clearly, we show the $Q_{\rm Kuiper}$ instead. So 0.1 corresponds to 90\% C.L. against uniformity of phases, 0.01 to 99\% C.L. \ldots etc..}  \label{cl}
\end{figure}

The peculiarity of $\l=3$ and 8 found by Copi \etal \shortcite{copi} can be seen clearly in the cross correlation
with the foregrounds at all 5 foreground maps, as shown in
Fig.\ref{cl}, particularly with the W band foreground map. This is
another advantage of using phases as
cross checking for non-Gaussianity for the CMB signal. There is no
other method so far that can cross check both the foreground maps and
the derived CMB map when peculiarities are found. What is unclear
though, is the non-random phases at $\l=6$ appearing on all ILC, EILC
WFM, and PCM (Fig.\ref{ksmapping}), where the cross correlation
with foregrounds is reasonably low.      


\section*{Acknowledgments}
We are grateful for the useful discussions with
Peter Coles and Mike Hobson. We acknowledge the use of
the Legacy Archive for Microwave Background Data Analysis
(LAMBDA). Support for LAMBDA is provided by the NASA Office of Space
Science. We thank Tegmark \etal and Eriksen \etal for providing their processed
maps. We also acknowledge the use of H{\sc
ealpix}\footnote{\tt http://www.eso.org/science/healpix/} package
\cite{healpix} to produce $\alm$ from the \wmap data.

\newcommand{\autetal}[2]{{#1\ #2. et al.,}}
\newcommand{\aut}[2]{{#1\ #2.,}}
\newcommand{\saut}[2]{{#1\ #2.,}}
\newcommand{\laut}[2]{{#1\ #2.,}}

%
%
\newcommand{\refs}[6]{#5, #2, #3, #4} 
\newcommand{\unrefs}[6]{#5, #2 #3 #4 (#6)}  

%
%
\newcommand{\book}[6]{#5, {\it #1}, #2} 
%
\newcommand{\proceeding}[6]{#5, in #3, #4, #2} 
%
%

\newcommand{\combib}[3]{\bibitem[\protect\citename{#1 }#2]{#3}} 

%
%
\def\apj{ApJ}
\def\apjl{ApJL}
\def\apjs{ApJS}
\def\mn{MNRAS}  
\def\nature{nat} 
\def\aa{A\&A}   
\def\prl{Phys.\ Rev.\ Lett.}
\def\prd{Phys.\ Rev.\ D}
\def\pr{Phys.\ Rep.}

\newcommand{\amp}{\& }

\def\cup{Cambridge University Press, Cambridge, UK} 
\def\princetonpress{Princeton University Press}
\def\worldpress{World Scientific, Singapore}
\def\oxfordpress{Oxford University Press}

\end{document}